\documentclass[]{article}
\usepackage[numbers,sort&compress]{natbib}
\usepackage[margin=1.25in]{geometry}
\usepackage{amsmath,amssymb,amsfonts}
\usepackage{algorithmic}
\usepackage{graphicx}
\usepackage{textcomp}
\usepackage{xcolor}

\usepackage[shortlabels]{enumitem}
\usepackage{tcolorbox}
\usepackage{amssymb}
\usepackage{amsmath}
\usepackage{amsthm}
\usepackage{mathtools}
\usepackage{hyperref}
\usepackage{float}

\usepackage{pgfplots}
\usepackage{pgfplotstable}
\usepackage{subcaption}

\usepackage[ruled]{algorithm2e}
\DeclareMathOperator*{\argmax}{argmax}

\DeclareMathOperator*{\BDPL}{BDPL}

\renewcommand{\epsilon}{\varepsilon}
\newcommand{\eps}{\varepsilon}
\newcommand{\bits}{\{0, 1\}}
\newtheorem{theorem}{Theorem}
\newtheorem{lemma}{Lemma}
\newtheorem{proposition}{Proposition}

\newtheorem{corollary}{Corollary}

\newtheorem{definition}{Definition}

\newcommand{\eqdef}{\overset{\mathrm{def}}{=}}

\hypersetup{
     colorlinks=true,      		% false: boxed links; true: colored links
     linkcolor=blue,        % color of internal links
     citecolor=magenta,     % color of links to bibliography
     filecolor=green,      		% color of file links
     urlcolor=orange,           	% color of external links
}

\title{Optimal Local Bayesian Differential Privacy over Markov Chains}
\author{Darshan Chakrabarti, Jie Gao, Aditya Saraf, Grant Schoenebeck, Fang-Yi Yu}

\begin{document}

\maketitle

\begin{abstract}
    In the literature of data privacy, differential privacy is the most popular model. An algorithm is differentially private if its outputs with and without any individual's data are indistinguishable. In this paper, 
    we focus on data generated from a Markov chain and argue that 
    Bayesian differential privacy (BDP) offers more meaningful guarantees in this context. Our main theoretical contribution is providing a mechanism for achieving BDP when data is drawn from a binary Markov chain. We improve on the state-of-the-art BDP mechanism and show that our mechanism provides the \emph{optimal} noise-privacy tradeoffs for any local mechanism up to negligible factors. We also briefly discuss a non-local mechanism which adds \emph{correlated} noise. Lastly, we perform experiments on synthetic data that detail when DP is insufficient, and experiments on real data to show that our privacy guarantees are robust to underlying distributions that are not simple Markov chains.
\end{abstract}

% \begin{CCSXML}
% <ccs2012>
%   <concept>
%       <concept_id>10003752.10010070.10010111.10011735</concept_id>
%       <concept_desc>Theory of computation~Theory of database privacy and security</concept_desc>
%       <concept_significance>500</concept_significance>
%       </concept>
%   <concept>
%       <concept_id>10002978.10003018.10003019</concept_id>
%       <concept_desc>Security and privacy~Data anonymization and sanitization</concept_desc>
%       <concept_significance>500</concept_significance>
%       </concept>
%  </ccs2012>
% \end{CCSXML}

% \ccsdesc[500]{Theory of computation~Theory of database privacy and security}
% \ccsdesc[500]{Security and privacy~Data anonymization and sanitization}

% \keywords{differential privacy, local privacy, correlated data, Markov chains}

\section{Introduction}
In the literature of privacy, differential privacy (DP)~\cite{dwork2014algorithmic} is the most popular model. An algorithm is differentially private if its outputs with and without any individual's data are (nearly) indistinguishable. It guards against what an adversary learns about an agent \textit{due directly} to that agent's participation. But this does not offer guarantees on how much an adversary can infer about personal data from the perturbed output.
Specifically, with correlated data, DP's guarantee is insufficient, as understanding the correlation can bolster inference significantly~\cite{Kifer2011no}. Real world data exhibits several natural correlation structures.  Social networks mediate interactions and influence which often lead to strongly correlated personal attributes.   Similarly, spatial and temporal proximity lead to strong correlations in data from sensors recorded as discretized time series. Some examples include data from human mobility traces, power grids, health data from personal wearable devices, and US census data.
In many applications the correlation structure can be learned from historical data and so should be assumed to be public knowledge. Let the \textit{correlated advantage} be how much an adversary can infer about private data from the perturbed output, given explicit knowledge of the correlation structure. We want a privacy definition and mechanism that bounds the correlated advantage.  

To incorporate correlated data, the Pufferfish framework, proposed by \citet{kifer2014pufferfish}, allows for robust privacy guarantees based on a specified data generation model and secret pairs.
Bayesian differential privacy (BDP), initially proposed by \citet{yang2015bayesian}, is an instantiation of the Pufferfish privacy framework and a strict generalization of differential privacy. The distinction between DP and BDP is most salient when comparing which adversaries they protect against. DP protects only against adversaries who know all but one tuple, while BDP simply quantifies over adversaries who know arbitrarily many tuples. As an example, suppose we are trying to protect Alice's time series data. Differential privacy guarantees \textit{only} that someone who knows Alice's location at all but one time step will not learn very much by analyzing the sanitized data set (their posterior after seeing the sanitized data will not change much)~\cite{yang2015bayesian}. BDP guarantees that, regardless of what the adversary knows about Alice's mobility data, they will not learn much by analyzing the sanitized data set.\footnote{Notice that when the data at each time step is independent, BDP reduces to DP; the adversary's knowledge of Alice's location at time $x_i$ is irrelevant to their knowledge of Alice's location at time $x_j$.} Presumably, Alice is not interested in protecting her data only against an adversary who knows almost all of it. Analogously, our privacy guarantee should not depend on how many data points the adversary knows.
\subsection{Our Contribution}

Our main theoretical contribution is providing a mechanism for achieving BDP when the data is drawn from a binary Markov chain.  We focus on \textit{non-interactive} mechanisms, which return ``sanitized'' (i.e. noisy) estimates of the real database. These are called non-interactive since the sanitized databases can be queried offline without further privacy loss; in contrast, query-based mechanisms must be rerun with additional privacy losses for each query. In order to achieve \textit{local} privacy guarantees, our primary mechanism adds independent noise to each tuple. Local privacy means that no centralized curator needs to have access to the agents' true data. Instead, the owners of each tuple can privately sanitize their entry before submission, which would be ideal for IoT settings. When privacy guarantees are framed in terms of trust or persuasive power, this property is extremely attractive.\footnote{Though we mostly focus on the setting where every tuple belongs to one individual, our mechanism is still applicable to the conventional setting.} 

Our model is simple yet fundamental; we represent data correlations via Markov chains, and these preliminary results consider only binary state spaces. We assume that the data is \textit{positively} correlated, as in the case of location data, where an individual is more likely to stay in the same location than leave (given a fine-enough time scale). We improve on the state-of-the-art BDP mechanism and show that our mechanism provides the \emph{optimal} noise-privacy tradeoffs (among local/independent noise mechanisms) up to negligible factors. This is significant because the previous general results only provide a sufficient bound on the noise~\cite{liu2016dependence, zhao2017dependent}. The main challenge with finding an exact bound is describing how the privacy loss evolves through the Markov chain. We also consider a mechanism which adds \textit{correlated} noise to the data, but find no additional improvement to noise-privacy tradeoffs. Lastly, we perform experiments on real and synthetic data. 
We first demonstrate that DP does not bound the correlated advantage, by providing a concrete, correlation aware attack that more than doubles the DP bound.
Further, even on real data that is not entirely Markovian, a neural network adversary, using Long Short Term Memory (LSTM) models, cannot surpass our mechanism's privacy bounds, suggesting that our mechanism is \textit{robust} to varying correlation structures in practice.

\subsection{Related Work}
There exists a large body of work regarding the limitation of differential privacy with correlated data.\footnote{See \cite{Kifer2011no, chen2014correlated, yang2015bayesian, song2017pufferfish,zhao2017dependent,he2014blowfish,zhu2014correlated} for examples in various contexts.} As a result, there have been many new correlation-aware privacy definitions in recent years. \citet{zhao2017dependent} consider a definition equivalent to BDP that they term \emph{dependent differential privacy}. Their proofs show that many attractive properties of DP, like post-processing and composition guarantees, also hold for BDP.  \citet{liu2016dependence} consider a different privacy definition that they also term dependent differential privacy; however, their definition is quite distinct, as it does not imply DP \cite{zhao2017dependent}.  \citet{naim2019off} consider a privacy definition rooted in information theory, that is more applicable to Internet privacy. 

As far as privacy preserving mechanisms, the original BDP paper considers a mechanism for the sum query, with data drawn from a Gaussian query model \cite{yang2015bayesian}. We model data generated from a Markov chain, so their mechanism does not apply.  \citet{song2017pufferfish} provide a very general mechanism along with guarantees for any privacy definition in the Pufferfish framework. However, their mechanism may be computationally intractable (e.g. it runs in $O(n^3)$ in our setting, while ours runs in time $O(n)$), and must be re-run (with additional privacy loss) for multiple queries. \citet{zhao2017dependent} provides a very general reduction theorem which explains how $\eps$-BDP is implied by a lower $\eps'$-DP. This result, while very powerful, does not produce optimal noise-privacy tradeoffs, as we explore in section 4. 

To the best of the authors' knowledge, this is the first paper with an explicit optimal mechanism for local, non-interactive privacy in correlated settings.

\smallskip\noindent\textbf{Roadmap} In section 2, we review some basic definitions, and clearly differentiate the semantic guarantees of DP and BDP. In section 3, we describe the noise-privacy tradeoffs of our mechanism, provide high level proof sketches, and briefly discuss a correlated noise variant of our mechanism. In section 4, we run several experiments and compare with prior results. We show that $\eps$-DP requires less noise than $\eps$-BDP, and also experimentally evaluate how much privacy degrades when an adversary knows the data correlation structures. We then compare our noise-privacy tradeoffs with previous results, showing that we improve on the state-of-the-art. To round out our experiments, we evaluate our mechanism on real world heart rate data which may not be purely Markovian, which shows that our mechanism is robust to real world correlation patterns.

\section{Preliminaries}
%To motivate our privacy framework, we use the example of heartbeat data. Let $X = X_1, \dots, X_n$ represent a time series of a person's heart rate, where $X_i$ corresponds to a discretization of their heart rate at the $i$th timestep. For our purposes, $X_i \in \bits$, so we consider $X_i$ as conveying whether their heart rate is elevated or not. The individual would like to publish their heart rate data as part of a medical study. However, they do not want anyone to be able to know, with high confidence, whether their heart rate is elevated at any given moment. 

Let $X = X_1, \dots, X_n$ represent a time series, where $X_i  \in \bits$ corresponds to the value at the $i^{\text{th}}$ timestep. The data curator would like to publish the time series but does not want anyone to be able to know, with high confidence, the exact value at any given moment. We now define the data generation model and provide formal privacy definitions for DP and BDP. Then, we explain the semantic difference between the two privacy definitions, and provide arguments in favor of BDP for our context. First, we define \emph{negligible} functions, which we use to explain how our mechanism is optimal.
\begin{definition}
A negligible function is a function $\mu: \mathbb{N} \to \mathbb{R}$ that is asymptotically bounded by every inverse polynomial.
\end{definition}
A common example is $2^{-n}$, an inverse exponential.
When we use negligible, we mean negligible in the length of the Markov chain.

\subsection{Markov Chains}
A sequence of random variables $(X_1, \ldots, X_n)$ is a \emph{Markov chain} with state space $\Omega$, initial distribution $\pi_0$, and transition matrix $P$ if 1) the initial state $X_1$ is sampled from $\pi_0$, and 2) for all $x, x'\in \Omega$, all $t\ge 1$, and all events $H_{t} = \{X_1 = x_1, \ldots, X_{t-1} = x_{t-1}, X_t = x\}$, we have 
$$\Pr[X_{t+1} = x'\mid H_t] = \Pr[X_{t+1} = x'\mid X_t = x] = P_{x, x'}$$
In this paper, we consider the special case where $\Omega = \bits$. We use $(\Omega, \pi_0, P)$ to denote a Markov chain. We call a Markov chain \emph{lazy} if $P(x,x) > 1/2$ for all $x\in \Omega$, meaning that the chain is more likely to stay in the current state than switch states.

A distribution $\pi$ on $\Omega$ is a \emph{stationary distribution} of the Markov chain $(\Omega, \pi_0, P)$ if $\pi = \pi P$. Thus, if $X_1$ is drawn from $\pi$, the marginal distribution of any state $X_i$ is also given by $\pi$.

\subsection{Differential Privacy}
A randomized mechanism $\mathcal{M}$ is a function with domain $\mathcal{X}$ consisting of all possible input databases, and range $S$ denoting all possible outputs. 
% such that the output $\mathcal{M}(X \in \mathcal{X}) = z \in S$ obeys a probability distribution conditioned on $X$: $\Pr(\mathcal{M}(X \in \mathcal{X}) = z \in S)$. 
%\end{definition}
\begin{definition}[\citet{dwork2014algorithmic}]
A randomized mechanism $\mathcal{M}$ is said to be \emph{$\eps$-differentially private} ($\eps$-\emph{DP}) if, for any databases $x$ and $y$ that differ in exactly one tuple (i.e. one data point),
\begin{align*}
    \sup_{s \in S} \frac{\Pr[\mathcal{M}(x) = s]}{\Pr[\mathcal{M}(y) = s]} \le e^\eps.
\end{align*}
\end{definition}
Databases $x$ and $y$ are said to be \emph{neighbors} and $\eps$ is referred to as the privacy budget.
Note that lower $\eps$ entails more privacy. 

\subsection{Bayesian Differential Privacy}
For BDP, we now assume that the domain $\mathcal{X}$ of $\mathcal{M}$ is generated according to some probabilistic model; thus, it makes sense to consider, for $X \in \mathcal{X}$, $\Pr[X]$ or $\Pr[M(X) = s|X_i = x_i]$. We also explicitly introduce adversaries $A = A(i, K)$, where $i \in [n]$ denotes the tuple $A$ is trying to infer and $K \subseteq [n] \setminus \{i\}$ denotes the tuples $A$ already knows.
\begin{definition}
The Bayesian Differential Privacy Loss (BDPL) of $\mathcal{M}$ with respect to $A$ is defined as:
\begin{align*}
    \BDPL(A; \mathcal{M}) \eqdef \sup_{x_i, x'_i,\mathbf{x}_K, s} \frac{\Pr[\mathcal{M}(X) = s|x_i, \mathbf{x}_K]}{\Pr[\mathcal{M}(X) = s|x'_i, \mathbf{x}_K]}
\end{align*}
\end{definition}
In the above definition, $x_i$ and $x'_i$ correspond to two different values for tuple $X_i$. Also, note that the probability is taken over the mechanism and data generation process. It should be interpreted as the probability of the database being $X$, given $x_i$ and $\mathbf{x}_K$, and then observing $\mathcal{M}(X) = s$. 
\begin{definition}
A randomized mechanism $\mathcal{M}$ is said to be $\eps$-Bayesian differential private ($\eps$-BDP)~\cite{yang2015bayesian}, if
\begin{align*}
\sup_{A} \BDPL(A; \mathcal{M}) \le e^\eps
\end{align*} 
where the adversaries range over all possible $i$ and $K$.
\end{definition}
$\eps$-DP is equivalent to requiring $\BDPL(A;\mathcal{M}) \le e^\eps$, for all adversaries $A = A(i, [n]\setminus \{i\})$, i.e all adversaries that know all but one tuple. It follows that $\eps$-BDP is at least as strong as $\eps$-DP.
% The only difference between the definition of $\eps$-DP and $\eps$-BDP is that in $\eps$-DP the expressions are conditioned on all the tuples besides the one that is being altered (at index $i$) which means that it is assumed that the adversary has knowledge of all of the other tuples, while in $\eps$-BDP, $\mathcal{K} \subseteq [n]\setminus \{i\}$ which means that the adversary's knowledge is flexible. It follows that because $\mathcal{K}$ could be equal to $[n] \setminus \{i\}$, $\eps$-BDP is at least as strong as $\eps$-DP (the associated BDP loss for a mechanism is at least as large as the DP loss for the same mechanism).

\subsection{Semantic Differences between DP and BDP}
Differential privacy forms privacy guarantees \emph{without} a model for how the data is generated. However, if an adversary has reasonable background knowledge regarding the data distributions, it may be the case that an $\eps$-differentially private mechanism will produce an output that is disproportionately more likely (from the perspective of the adversary) given one of two neighboring datasets. This limitation of DP is well understood;~\citet{dwork2014algorithmic} reframes this limitation in terms of the \emph{semantic} guarantee that DP provides. Suppose you are a medical researcher tasked with convincing users to divulge sensitive health data, which will then be published online. Differential privacy, according to the semantic interpretation, can be used as a \emph{tool} to encourage participation. Individuals, who can only control their own participation in the study, know that they will receive minimal (privacy) harms \emph{directly tied} to their participation in the study. Viewed this way, DP is an entirely end-user focused persuasive tool. However, an ethical researcher may not only like to persuade users to participate, but also to understand and limit the harms caused by the study itself. DP answers the question of whether a single user ought to participate; BDP answers the question of whether the study ought to be performed (in terms of the privacy ``cost'' of the study). Put another way, BDP persuades the researcher to publish a sanitized copy of their data, by more comprehensively limiting the harms to any study participant.

We stress that in this medical example, since the database consists of records that each belong to different individuals, DP can still be a reasonable choice (e.g. if the study could significantly help the participants, and the additional noise hinders utility). However, when considering data from a single individual, like mobility or heartbeat time series data, (standard) DP does not provide sufficient guarantees.\footnote{There are definitions such as group differential privacy~\cite{dwork2014algorithmic} that apply to this setting, but they require significantly more noise. See~\cite{song2017pufferfish} for a more thorough analysis.} 
%Since our central example and experimental data both concern heartbeat time series data from single individuals, 
BDP provides much more meaningful guarantees than DP in these settings.

\section{Optimal Mechanisms for Bayesian Differential Privacy}

In this section, we introduce our main theoretical results and some high-level proof sketches. We consider a variant of the canonical randomized response mechanism, applied to data with correlations modeled by a Markov chain. 

Suppose we have a dataset $X=(X_1, \dots, X_n)$ generated from a Markov chain on $\bits$. For example, this data could be discretized heart rate data, indicating whether a person's heart rate is currently elevated or not.
% Notice that such data is clearly not independent (on a fine timescale), since if someone is walking/running right now, they are more likely to continue their current activity. 
We assume the Markov chain is stationary and has transition matrix:
\begin{align*}
    P = \begin{bmatrix}
    1 - q & q \\
    r & 1 - r
    \end{bmatrix}, \text{ where } q, r \in (0, 0.5)
\end{align*}
The initial state is generated from the stationary distribution, $\pi = \left(\frac{r}{q+r}, \frac{q}{q+r}\right)$, so the marginals are given by $\pi$. Such a Markov chain is called \emph{lazy}, since we are more likely to stay in the current state than transition. Our proposed mechanism, \autoref{mainMech}, takes in database $X$, adds independent noise to each data point, and outputs sanitized database $Z$. 
\begin{algorithm}
    \KwIn{A sequence of data $(x_1, \ldots, x_n)$ sampled from Markov chain $(\Omega, \pi, P)$, and noise levels $\rho_0, \rho_1 \in [0,1/2)$}
    \KwResult{Sanitized data $(z_1, \ldots, z_n)$.}
    \For{$i = 1,\ldots, n$}{
    Generate $z_i$ such that $\Pr[Z_i = z\mid X_i = x] = B_{x,z}$ independently, where 
    \begin{equation*}
        B = \begin{bmatrix}
    1 - \rho_0& \rho_0\\
    \rho_1 & 1 - \rho_1
    \end{bmatrix}
    \end{equation*}
    }
    \caption{Independent Noise Mechanism}
    \label{mainMech}
\end{algorithm}
We perturb a $0$ state to a $1$ state with probability $\rho_0$ and a $1$ state to $0$ with probability $\rho_1$. We use two separate noise levels to account for cases where one state is much more likely than the other; we can add less noise to the more popular state to reduce the total expected noise significantly. Notice also that this mechanism can be implemented in a distributed fashion, to guarantee local privacy. We want this mechanism to satisfy $\eps$-BDP, and our goal is to determine $\eps$ in terms of parameters $\rho_0, \rho_1, q$ and $r$.

BDP requires us to protect against attackers who might know any number of tuples. To simplify our work, we will show later that the most ignorant adversary requires the most noise to protect against.\footnote{This in turn implies that DP, which assumed a fully informed adversary, will not provide a strong privacy guarantee.} This result, while somewhat counterintuitive, matches other cases with positive data correlations~\cite{yang2015bayesian}. So, by \autoref{thm:bdpProof}, $\eps$-BDP can be reduced to the following conditions:
\begin{align}
    \max_{i \in [n], \mathbf{z} \in \bits^n} \frac{\Pr(\mathbf{Z} = \mathbf{z}|X_i = 0)}{\Pr(\mathbf{Z} = \mathbf{z}|X_i = 1)} \le e^\eps \label{bdpDefTop} \\
    \max_{i \in [n], \mathbf{z} \in \bits^n} \frac{\Pr(\mathbf{Z} = \mathbf{z}|X_i = 1)}{\Pr(\mathbf{Z} = \mathbf{z}|X_i = 0)} \le e^\eps \label{bdpDefBot}
\end{align}

Where $\mathbf{Z} = \mathbf{z}$ denotes the event that $Z_1 = z_1, \dots, Z_n = z_n$. When it is clear from context, we will sometimes use $\Pr[z_{i:j}|x_k]$ as a short form of $\Pr[Z_{i:j} = z_{i:j}|X_k = x_k]$. 
% Since we want to solve for $\eps$ in terms of $\theta$ and $\rho$, we'll need to determine which values of $j$ and strings $z$ maximize/minimize this probability ratio. 
We will often refer to \autoref{bdpDefTop} and \autoref{bdpDefBot} as \emph{likelihood ratios}, as they measure the probability of the data given competing hypotheses $X_i = 0$ and $X_i = 1$.

We can view the original and sanitized data as a Hidden Markov Model (HMM), with hidden states $X$, observed states $Z$, transition matrix $P$ and emission matrix $B$. We are interested in computing $\Pr[\mathbf{z}|x_j]$. Given concrete $q, r, \rho_0, \rho_1$, this can be solved via the Forwards-Backwards algorithm, but we are interested in providing a closed form with these as variables. The theorem below is our main technical result, providing a tight, closed form bound.
% The adversary sees the observations and should not be able to guess the hidden states with high probability. The transition probabilities are represented by the matrix $P$ from above, and the emission probabilities by the noise matrix $B$ from \autoref{mainMech}. We are interested in the quantity $\Pr[z | x_j = b]$. Solving for $\Pr[x_j = b| z]$ is known as \emph{smoothing} in the HMM literature. The Forwards-Backwards algorithm is typically used to solve this problem, but the typical descriptions assume that the probability distributions are known. In our case, $P$ and $B$ are parameterized distributions, and we're looking for a closed form of $\Pr[z | x_j = b]$ in terms of $\rho_0, \rho_1, q$ and $r$ in order to find the pairs $(z, j)$ that maximize \autoref{bdpDefTop} and \autoref{bdpDefBot}. We'll focus on \autoref{bdpDefTop}, as \autoref{bdpDefBot} is symmetrical with our current model.  The theorem below summarizes our main results for this section.

\begin{theorem} \label{thm:maxLR}
    Let $0 < q, r, \rho_0, \rho_1 < 0.5$. For all $n \ge 1$, all $\mathbf{z} \in \bits^n$, and all $i \in [n]$,
    \begin{align*}
        &\frac{\Pr[\mathbf{Z} = \mathbf{z}\mid X_i = 0]}{\Pr[\mathbf{Z}= \mathbf{z}\mid X_i = 1]} \le \frac{a^2}{cd}  \text{, where} \\
        a &= \big((1 - q)^2(1 -\rho_0)^2 - 2(1 - q - r - qr)(1 -\rho_0)\rho_1 \\
        &\quad +(1 - r)^2\rho_1^2\big)^\frac{1}{2} + (1 -\rho_0)(1 - q) - \rho_1(1 - r)\\
        c &= 2r\rho_1, \quad d = 2r(1-\rho_0).
    \end{align*}
    Moreover, the maximum happens at $\mathbf{z} = \mathbf{0}$ and $i \approx n/2$. With these parameters, the bound is tight up to negligible factors (in the length of the Markov chain).
\end{theorem}

\begin{figure}%[h!]
    \centering
    \begin{subfigure}{.45\columnwidth}
        \centering
        \includegraphics[width=0.9\linewidth,keepaspectratio]{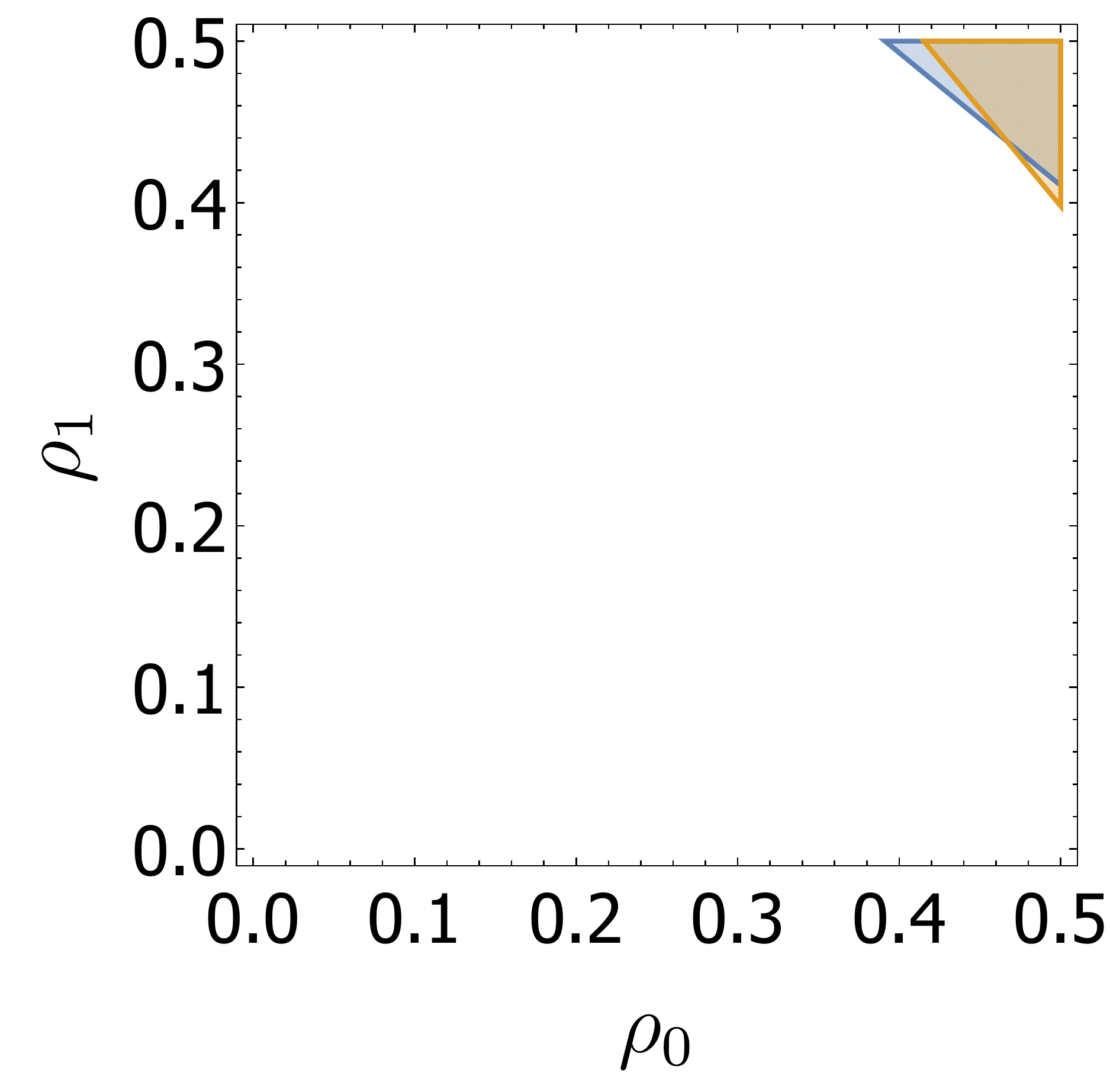}
        \subcaption{}
    \end{subfigure}
    \begin{subfigure}{.45\columnwidth}
        \centering
        \includegraphics[width=0.9\linewidth,keepaspectratio]{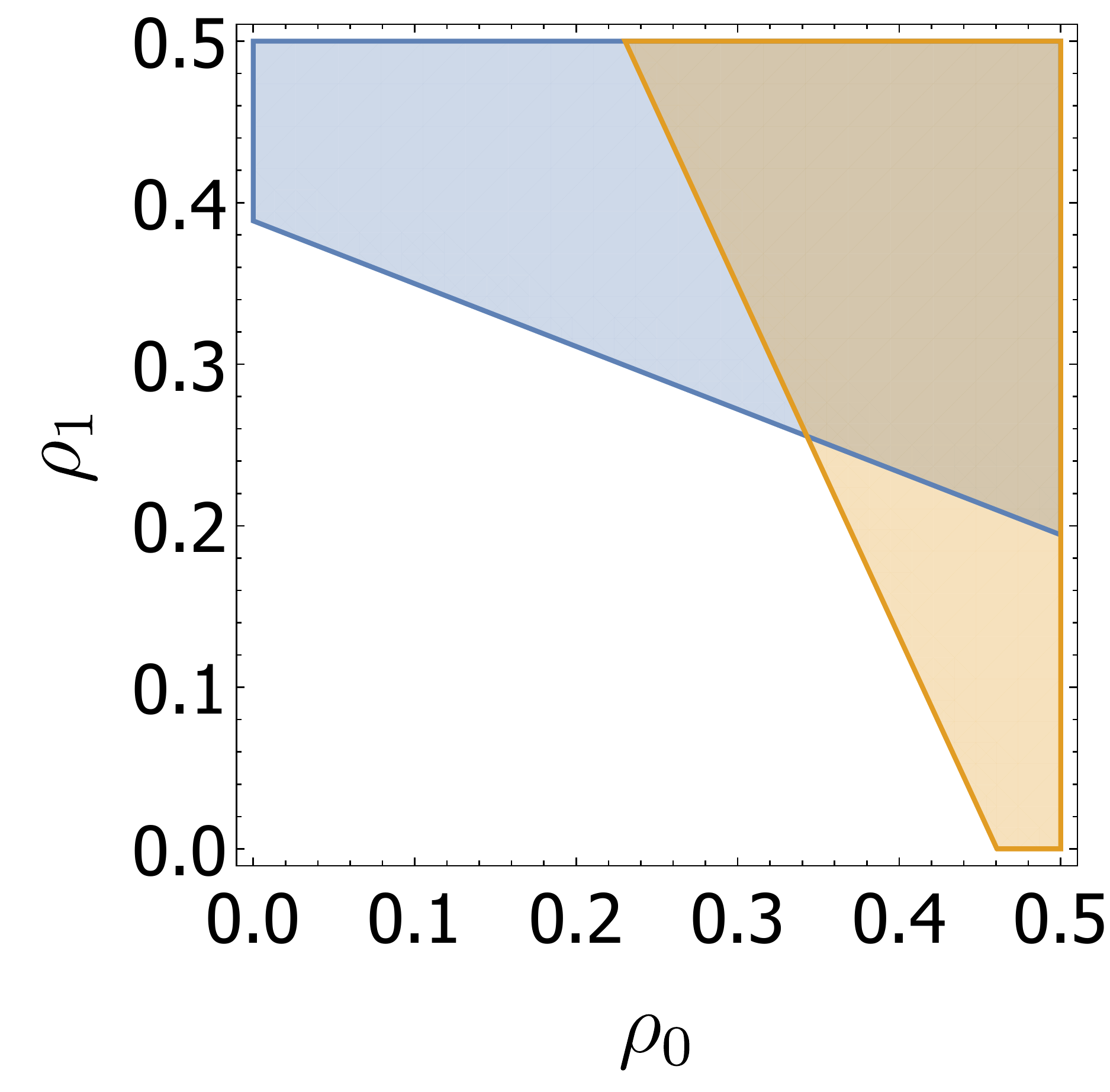}
        \subcaption{}
    \end{subfigure}
    \caption{Noise Constraints for BDP, on a Markov chain with $q = 0.2$, $r = 0.35$, and $\eps = 0.5$ (a) or $\eps = 2$ (b). Each plane represents one of the constraints; the intersection represents the feasible set of noise parameters.}
    \label{fig:noise_regions}
\end{figure}

The theorem says that we achieve (nearly) optimal noise-privacy tradeoffs, for our data model and independent noise algorithm. By symmetry, the likelihood ratio for \autoref{bdpDefBot} simply swaps $q$ with $r$ and $\rho_0$ with $\rho_1$ in the theorem.
Notice that there are infinitely many solutions for $\rho_0$ and $\rho_1$ for fixed $\eps, q, r$. In \autoref{fig:noise_regions} we plot the regions of feasible noise parameters for a fixed Markov chain (each region corresponds to one of the BDP constraints).
The figure suggests a linear relationship between $\rho_0$ and $\rho_1$. Indeed, this can be rigorously shown, and implies that any linear objective function (like the expected noise) can be minimized by looking only at a constant number of points. We omit the exact forms of these linear constraints in this version.
\begin{corollary}
    For a fixed privacy budget $\eps$ and known Markov chain parameters $q, r$, to minimize the expected noise (i.e. $\rho_0 \cdot \Pr[X_i = 0] + \rho_1 \cdot \Pr[X_i = 1]$), one can determine the optimal $\rho_0$ and $\rho_1$ in constant time, up to negligible error in $n$.
\end{corollary}
For many of our comparative discussions, we focus on the special case of a symmetric Markov chain, where the states have identical transition probabilities and noise levels. We get the following simplified results:
\begin{corollary} \label{cor:symMC}
    Assume the Markov chain is symmetric. Let $\theta = q = r$ and $\rho = \rho_0 = \rho_1$. For all $n\ge 1$, all $z \in \{0,1\}^n$, and all $i\in [n]$:
    % \footnotesize
    \begin{align*}
        &\frac{\Pr[z\mid X_i = 0]}{\Pr[z\mid X_i = 1]} \le \frac{1-\rho}{\rho} \left(\frac{a}{c}\right)^2 \text{, where} \\
        a &= \sqrt{\theta^2 + (1- 2\theta)(1-2\rho)^2} + (1-\theta)(1-2\rho)\\
        c &= 2\theta(1-\rho)
    \end{align*}
    \normalsize
    Moreover, given $\theta, \rho$ and $n$, the maximum happens at $\mathbf{z} = \mathbf{0}$ and $i = \lfloor n/2\rfloor$. Further, given $\theta$, for a desired privacy level $\eps$, it suffices to set:
    \begin{align*}
	\rho \ge \frac{4 + \theta(\theta e^\eps -2) - \sqrt{\theta^2 e^\eps(4+\theta(\theta e^\eps-4))}}{8+2\theta(\theta e^\eps + \theta - 4)}
    \end{align*}
\end{corollary}
This closed form is not exact; to see how much this approximation loses, see \autoref{fig:thm3_ddp_noise}.

\subsection{Proof Sketch for \autoref{thm:maxLR}}
We prove \autoref{thm:maxLR} in four parts. First, we write the likelihood ratio (LR) in terms of modified $\alpha$, $\beta$ recurrences for HMMs. Next, we show that for any fixed index $i$, the value of $\mathbf{z}$ that maximizes the LR is $\mathbf{z}=\mathbf{0}$, i.e. the all-zero database. Then, we find a closed form for the $\alpha$ and $\beta$ recurrences when $\mathbf{z}=\mathbf{0}$ via matrix diagonalization. This closed form allows us to prove the bound in the theorem. Lastly, we show that for $\mathbf{z} = \mathbf{0}$, the LR is maximized with $i \approx n/2$, and that our bound is tight up to inverse exponential factors. We adopt this approach because a closed form of the LR in terms of the parameters is hard to compute for arbitrary $\mathbf{z}$.

Because the algorithm can be viewed as a hidden Markov model, the likelihood ratio can be computed by a slight variant of the standard Forwards-Backwards decomposition \cite{russell_norvig_2020}. Given an outcome $\mathbf{z}\in \{0,1\}^n$ we can define the forward probability from state $1\le t\le n$ as: 
\begin{equation}
    \alpha_t(x; \mathbf{z}) := \Pr[\mathbf{Z}_{1:t} = \mathbf{z}_{1:t}\mid X_t = x]
\end{equation}  
Similarly we define the backward probability as:
\begin{equation}
    \beta_t(x; \mathbf{z}) := \Pr[\mathbf{Z}_{t+1:n} = \mathbf{z}_{t+1:n}\mid X_t = x]
\end{equation}
To simplify notation, we sometimes omit the outcome $\mathbf{z}$ when there is no ambiguity.
With the forward and backward probability, we can compute the likelihood ratio easily:
$$\frac{\Pr[\mathbf{Z} = \mathbf{z}\mid X_i = 0]}{\Pr[\mathbf{Z}= \mathbf{z}\mid X_i = 1]} = \frac{\alpha_i(0; \mathbf{z})}{\alpha_i(1; \mathbf{z})}\cdot \frac{\beta_i(0; \mathbf{z})}{\beta_i(1; \mathbf{z})}$$
Therefore, to maximize the likelihood ratio, it suffices to maximize the above $\alpha$ and $\beta$ ratios independently. The $\alpha$ and $\beta$ terms can be written recursively as\footnote{The base cases are $\alpha_0(x;\mathbf{z}) = 1$ and  $\beta_n(x; \mathbf{z}) = 1$ for all $x, \mathbf{z}$.}:
\begin{align*}
    \alpha_{i+1}(x;\mathbf{z}) &= \alpha_i(0)\,B_{x, z_{i+1}}\, P_{x, 0} +\alpha_i(1)\,B_{x, z_{i+1}}\, P_{x, 1} \\
    \beta_i(x; \mathbf{z}) &= \beta_{i+1}(0) B_{0, z_{i+1}} P_{x,0} + \beta_{i+1}(1)B_{1,z_{i+1}}P_{x,1}
\end{align*}
where $P$ and $B$ are the transition and noise matrices. 
Thus, to compute a closed form of the ratios, we must work with a concrete $\mathbf{z}$.   
\begin{lemma} \label{lm:lrMaxZ}
    For all $i\in [n]$, the likelihood ratio is maximized when $\mathbf{z} = \mathbf{0}$.  That is, for all $\mathbf{z} \in \bits^n$:
    $$\frac{\Pr[\mathbf{Z} = \mathbf{z} | X_i = 0]}{\Pr[\mathbf{Z} = \mathbf{z} | X_i = 1]} \le \frac{\Pr[\mathbf{Z} = \mathbf{0} | X_i = 0]}{\Pr[\mathbf{Z} = \mathbf{0} | X_i = 1]}.$$
\end{lemma}
The proof can be found in Appendix \ref{appendix1}. 
Since we are interested in computing the maximum likelihood ratio, over all $\mathbf{z}$ and $i$, the above lemma shows that we can restrict attention to $\mathbf{z} = 0$. To compute the theorem upper bound, we use the following lemma:
\begin{lemma}\label{lm:closedForm}
    With $\mathbf{z} = \mathbf{0}$, for all $i \in [n]$:
    \begin{align*}
    \frac{\alpha_{i}(0;\mathbf{z})}{\alpha_{i}(1;\mathbf{z})} \le \frac{a}{c} \mbox{ and } \frac{\beta_{i}(0;\mathbf{z})}{\beta_{i}(1;\mathbf{z})} \le \frac{a}{d}
\end{align*}
\end{lemma}
The proof can be found in Appendix \ref{appendix2} and involves writing the $\alpha$ and $\beta$ recurrences as matrices and diagonalizing for a closed form. However, the above lemma does not indicate how tight this bound is. To conclude the proof of the theorem, we show that a tuple that's roughly in the middle of the Markov chain is the easiest to attack, and that at this tuple, the bound we provide on the likelihood ratio is very tight.
\begin{lemma} \label{lm:lrMaxI}
    $$ \argmax_{i \in [n]} \frac{\Pr[\mathbf{Z} = \mathbf{0}|X_i = 0]}{\Pr[\mathbf{Z} = \mathbf{0}|X_i = 1]} = \frac{n}{2} \pm \mathcal{O}(1)$$
    Where $\mathcal{O}(1)$ refers to some constant that depends on the Markov chain parameters and noise parameters, but not on $n$, the length of the Markov chain. Further, with $i = n/2$, the bound in \autoref{thm:maxLR} is tight up to inverse exponential factors.
\end{lemma}
The proof can be found in Appendix \ref{appendix3}. With this, we've shown that our mechanism provides the optimal noise-privacy tradeoffs for BDP in our setting, up to negligible factors.

\subsection{The Most Ignorant Adversary is Hardest to Protect Against}
In \autoref{thm:maxLR}, we consider the likelihood ratio with respect to the adversary who knows none of the tuples. In order to show that our mechanism satisfies BDP, we show that this most ignorant adversary is the hardest to protect against. Although this may seem counter-intuitive, we will show later than the noise required for DP (which protects against the most informed adversary) is significantly lower than the noise in \autoref{thm:maxLR}. Here, we will show that the BDPL increases monotonically between these two extremes. We do so by showing that for any adversary $A$, an adversary $A'$ who knows one fewer tuple cannot have lower privacy loss:
\begin{theorem}\label{thm:bdpProof}
    Let $A = A(i, K)$ be an arbitrary adversary and let $U = [n] \setminus (K \cup \{i\})$. $K$ represents the known tuples and $U$ the unknown tuples for $A$. Consider adversary $A' = A(i, K')$, where $K' = K \setminus \{j\}$, for arbitrary $j \in K$. Then:
    \begin{enumerate}[(a)]
        \item if $j$ is not adjacent to $U$ or $i$, $\BDPL(A') = \BDPL(A)$.
        \item if $j$ is adjacent to $U$ or $i$, $\BDPL(A') > \BDPL(A)$.
    \end{enumerate} 
    Further, the maximum $\BDPL(A')$ occurs with $\mathbf{z}, \mathbf{x}_K, x_i = \mathbf{0}$.
\end{theorem}

% \fang{In the proof, it seems like we assume symmetric Markov chain.  We should state that if that's the case.  Furthermore, ``we confirmed numerically that $h(k) \eqdef f(k)g(k)$ is increasing in $k$. This completes the proof.'' is a pretty fishy, because such statement depends on $k$, $\theta$, and $\rho$.  In particular, it $h$ is a constant when $\theta = 1/2$ and $\theta = 0$, so part (b) in the theorem is equality $\BDPL(A') = \BDPL(A)$.  Frankly, I am not sure how to address this:  1) remove this result 2) show this result is true given $h$ is increasing 3) change (b) to $\BDPL(A') \ge \BDPL(A)$ and hope it is correct.}
In particular, this theorem implies that as the set $U$ of unknown tuples grows, in order for each unknown tuple to impact (and increase) the likelihood ratio, the unknown tuples must form a contiguous set around tuple $x_i$. The adversary's uncertainty about $x_i$ ``spreads'' and is amplified by $U$; the amount that $\mathbf{z}$ reveals is thus maximized when $U = [n] \setminus \{i\}$. The proof is fairly involved, but follows the same basic structure as the proof of \autoref{thm:maxLR}. It's obvious that the BDPL will always be maximized with $x_i, \mathbf{x}_{K}, \mathbf{z} = \mathbf{0}$, since the Markov chain is lazy and the mechanism is more likely to keep a bit than flip it (this can be shown formally via induction, similar to \autoref{lm:lrMaxZ}). Note that via conditional independence we can write:
\begin{align*}
    \BDPL(A) &= \frac{\Pr[\mathbf{z}|x_i, \mathbf{x}_K]}{\Pr[\mathbf{z}|\bar{x}_i, \mathbf{x}_K]} \\
    &= \frac{\Pr[z_i|x_i]}{\Pr[z_i|\bar{x}_i]}\cdot\frac{\Pr[\mathbf{z}_U|x_i, x_K]}{\Pr[\mathbf{z}_U|\bar{x}_i, x_K]}
\end{align*}
We then come up with a recurrence for the second term and find a closed form. With this, we show that BDPL of an adversary with $k$ unknown tuples contiguous with $x_i$ is proportional to a function $h(k)$. Lastly, we numerically verify that $h(k)$ is increasing in $k$ to complete the proof. The full proof can be found in Appendix \ref{appendix4}. For simplicity, we present the proof for the symmetric case, but the result holds in the asymmetric case as well. 

\subsection{Can Correlated Noise Help?}
We briefly report results on a mechanism which uses \emph{correlated} noise rather than independent noise to hide the data.
We restrict attention to data $\mathbf{x} \in \bits^n$ generated from a \emph{symmetric}, stationary, lazy Markov chain. The noise $\mathbf{y} \in \bits^n$ is also generated from a stationary Markov chain with parameters:
\begin{align*}
    B = \begin{bmatrix}
    1 - \rho_0 & \rho_0 \\
    \rho_1 & 1 - \rho_1
    \end{bmatrix}&\text{, where } \rho_0 < \rho_1 \in (0, 1) \\ &\text{ and }\pi_B = \left(\frac{\rho_1}{\rho_0+\rho_1}, \frac{\rho_0}{\rho_0+\rho_1}\right)
\end{align*}
The mechanism constructs the sanitized database by XORing the noise chain and data chain elementwise. This mechanism includes independent noise as a special case (when $\rho_0 + \rho_1 = 1)$. Notice that the expected noise for any bit is $\frac{\rho_0}{\rho_0+\rho_1}$, so if $\rho_0 \ge \rho_1$, then our mechanism is more likely to flip a bit than not. So, without loss of generality, we assume that $\rho_0 < \rho_1$ -- each bit is more likely to be preserved than flipped. 

With the further assumptions that $\rho_0 + \rho_1 \le 1$ and $\theta < \rho_0$ (made for mathematical convenience), we computed the optimal noise for this mechanism (up to negliglible factors). However, we found that the average noise was the same as our independent noise mechanism.\footnote{We found a closed form for $\rho_0$ and $\rho_1$ and computationally verified the expected noise equivalence. Proofs are omitted in this version.} This shows that our independent noise mechanism has the optimal noise-privacy tradeoffs over not just all independent noise mechanisms, but also over this subset of correlated noise mechanisms. 

\section{Comparative and Experimental Evaluations}
We first establish the necessity of a correlation-aware privacy definition, by demonstrating the efficacy of a concrete reconstruction attack on synthetic data that is protected by a differentially privacy mechanism. We then illustrate how our mechanism outperforms the state-of-the-art on BDP. Lastly, we perform experiments on Markov-like real data to demonstrate that a neural network (with a Long Short Term Memory architecture) cannot exploit any additional correlation structures to breach our BDP guarantees.
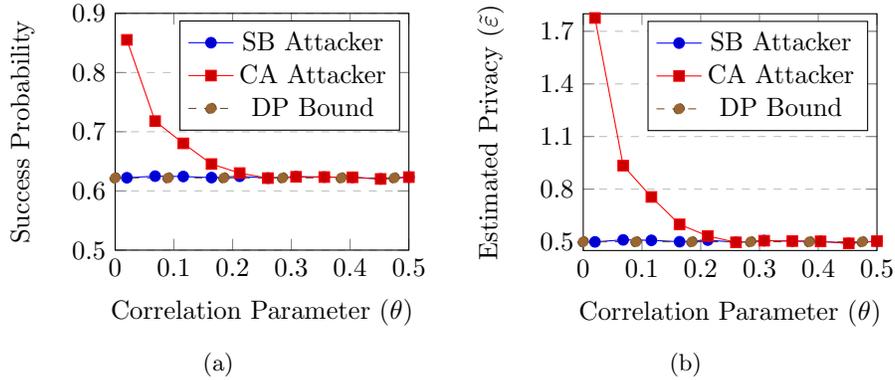
\begin{figure*}[t]
    \centering
    \begin{subfigure}{.4\textwidth}
        \centering
        \begin{tikzpicture}
            \begin{axis}[
                xlabel={Correlation Parameter ($\theta$)},
                ylabel={Success Probability},
                table/col sep=comma,
                xmin=0, xmax=0.5,
                ymin=0.5, ymax=0.9,
                xtick distance={0.1},
                ytick distance={0.1},
                legend pos=north east,
                ymajorgrids=true,
                grid style=dashed,
                width=0.9\linewidth,
            ]
            \addplot table [x=theta, y=suc_DP]{"data/eps0.5_suc.csv"};
            \addlegendentry{SB Attacker}
            \addplot table [x=theta, y=suc_BDP]{"data/eps0.5_suc.csv"};
            \addlegendentry{CA Attacker} 
            \addplot+[mark=*, dashed] coordinates{(0.0, 0.622) (0.09, 0.622) (0.185, 0.622) (0.285, 0.622) (0.385, 0.622) (0.475, 0.622)};
            \addlegendentry{DP Bound}
            \end{axis}
        \end{tikzpicture}
        \subcaption{}
    \end{subfigure}
    \begin{subfigure}{.4\textwidth}
        \centering
        \begin{tikzpicture}
            \begin{axis}[
                xlabel={Correlation Parameter ($\theta$)},
                ylabel={Estimated Privacy ($\tilde{\eps}$)},
                table/col sep=comma,
                xmin=0, xmax=0.5,
                ymin=0.45, ymax=1.8,
                xtick distance={0.1},
                ytick={0.5, 0.8, 1.1, 1.4, 1.7, 2.0},
                legend pos=north east,
                ymajorgrids=true,
                grid style=dashed,
                width=0.9\linewidth,
            ]            
            \addplot table [x=theta, y=eps_DP]{"data/eps0.5_eps.csv"};
            \addlegendentry{SB Attacker}
            \addplot table [x=theta, y=eps_BDP]{"data/eps0.5_eps.csv"};
            \addlegendentry{CA Attacker}
            \addplot+[mark=*, dashed] coordinates{(0.0, 0.5) (0.09, 0.5) (0.185, 0.5) (0.285, 0.5) (0.385, 0.5) (0.475, 0.5)};
            \addlegendentry{DP Bound}   
            \end{axis}
        \end{tikzpicture}
        \subcaption{}
    \end{subfigure}
    \caption{Experimental results confirm that the Single Bit (SB) attacker is bounded by DP. However, the Correlation Aware (CA) attacker can dramatically violate the DP bounds on highly correlated data. The probability of successfully recovering a hidden bit is plotted in (a) and the estimated privacy costs of the reconstruction probabilities are plotted in (b). Here, the true $\eps = 0.5$, and we generate data from a symmetric Markov chain, with transition parameter $\theta$. Lower $\theta$ means more correlation.}
    \label{fig:suvs}
\end{figure*}
\subsection{DP is Insufficient Against Correlated Advantage}
We first analyze the relationship between $\eps$ and $\rho$ for $\eps$-DP. Our mechanism $\mathcal{M}$ achieves $\eps$-DP if for all $\mathbf{z} \in \bits^n$ and all neighboring databases $\mathbf{x}, \mathbf{y} \in \bits^n$:
\begin{align*}
    \frac{\Pr[\mathcal{M}(\mathbf{x}) = \mathbf{z}]}{\Pr[\mathcal{M}(\mathbf{y}) = \mathbf{z}]} \le e^\eps
\end{align*}
Let $\mathbf{x}$ and $\mathbf{z}$ be arbitrary. Let $k$ be the number of indices that $\mathbf{x}$ and $\mathbf{z}$ agree in. Note that $\Pr[\mathcal{M}(\mathbf{x}) = \mathbf{z}] = (1-\rho)^k \rho^{n-k}$. Since $\mathbf{y}$ can differ in exactly 1 bit from $\mathbf{x}$:
\begin{align*}
    \frac{\Pr[\mathcal{M}(\mathbf{x}) = \mathbf{z}]}{\Pr[\mathcal{M}(\mathbf{y}) = \mathbf{z}]} \le \max\left(\frac{\rho}{1-\rho}, \frac{1-\rho}{\rho}\right) \le \frac{1 -\rho}{\rho}
\end{align*}
Thus, we get $\eps$-DP when:
\begin{align}\label{noiseForDP}
    \rho \ge \frac{1}{e^\eps + 1}
\end{align}
However, $\eps$-DP only considers the most knowledgeable attacker \footnote{In DP, the adversary need not know the correlation structure. However, when the adversary knows all but one data point, BDP and DP are equivalent.\cite{yang2015bayesian}}, and thus this noise level is insufficient to achieve $\eps$-BDP. This is somewhat counter intuitive -- one would expect that the most knowledgeable adversary is also the most difficult to protect against. However, one can also think about the most knowledgeable adversary as having the least uncertainty about $X_i$. If the adversary knows the values of $\mathbf{x}_{-i}$, then they can already use the Markov chain to compute an approximate distribution of $X_i$. In fact, the \emph{only} information they get from observing $\mathbf{z}$ is encoded in $\Pr[X_i | Z_i = z_i]$, since all $\mathbf{z}_{-i}$ are conditionally independent of $X_i$ given $X_{i-1}, X_{i+1}$. On the other hand, the adversary who knows none of the data points gets a lot more information from $\mathbf{z}$. Their prior over $X_i$ is simply $\pi$, but they could use the standard Forward-Backward inference algorithm to compute a much better informed posterior.

We also present some experimental results to highlight the practical ramifications of BDP over DP. First, consider an attacker with no information about the data generation. To attack variable $X_i$, the best the attacker can do is guess $z_i$ (since $\rho < 1/2$). Call this the Single Bit (SB) Attacker. On the other hand, a Correlation Aware (CA) attacker can run the Forward-Backward algorithm and output the most likely $x_i$, given $\mathbf{z}$. We ran a comparison of these two attackers on synthetic data. The data consists of length $30$ Markov chains. For a range of $\theta \in [0, 0.5]$, we generated $100$ different databases from a symmetric Markov chain with transition parameter $\theta$, and for each model, we generated $1000$ different sanitized databases. We then calculated the frequency of correctly guessing hidden state $X_{15}$ based on the sanitized database, for each type of attacker.\footnote{Our theoretical results show that the middle state maximizes the likelihood ratio, and so should be the easiest to distinguish.} 

To calibrate our expectations, let the success probability be $p_s$. For $\eps$ differential privacy, distinguishing the middle bit is the same as distinguishing $\mathcal{M}(x) = z$ from $\mathcal{M}(y) = z$, where $x$ and $y$ differ in the middle bit. So
\begin{align*}
    \frac{p_s}{1-p_s} \le \frac{\Pr[\mathcal{M}(x) = z]}{\Pr[\mathcal{M}(y) = z]} \le e^\eps \iff   p_s \le \frac{e^\eps}{1+e^\eps}
\end{align*}
Which, with $\eps = 0.5$, works out to roughly $p_s \le 0.622$. Our experimental results confirm that this bounds the SB attacker. However, the CA attacker beats this bound significantly with highly correlated data. In \autoref{fig:suvs}, we plot this and the privacy budget ``charged'' by each attacker, i.e. the $\eps$ that satisfies the previous equation for the observed $q$. Our results suggest that as the data are more correlated, we should add more noise to achieve the same privacy guarantees; this is exactly what we see with BDP, since $\rho$ is inversely proportional to $\theta$ in \autoref{thm:maxLR}.

\subsection{Comparison to Previous BDP Results}
\begin{figure}
    \includegraphics[scale=0.32]{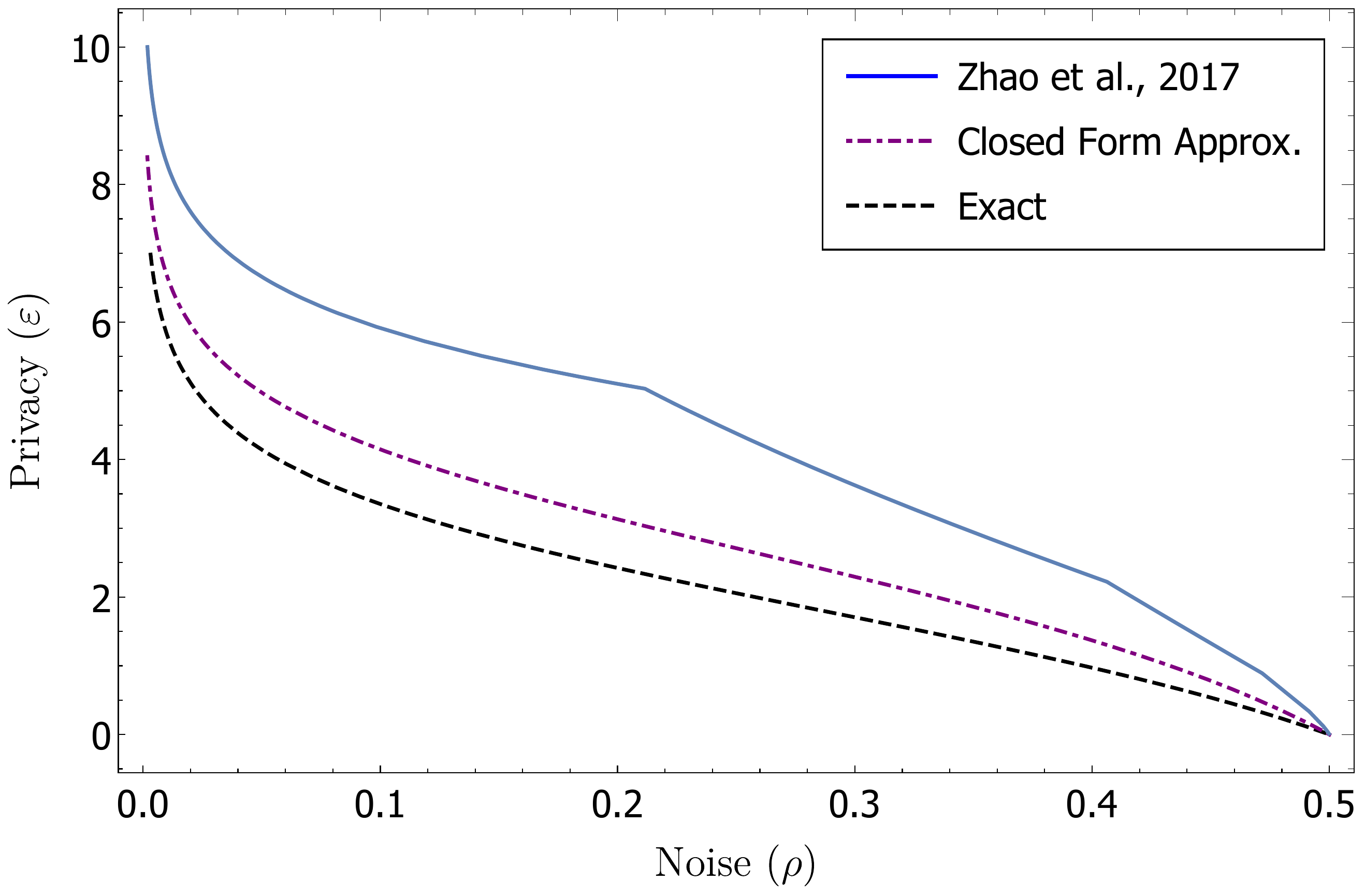}
    \caption{We plot the noise-privacy curves for a symmetric Markov chain with $\theta = 0.35$ and $n = 30$. The top line is the mechanism from \cite{zhao2017dependent}, the middle line our closed form approximation and the bottom line our exact bound in \autoref{cor:symMC}. Lower noise-privacy curves are better.}
    \label{fig:thm3_ddp_noise}
\end{figure}

Here, we briefly compare our results to that in \cite{zhao2017dependent} which uses a privacy definition equivalent to BDP. They provide very general theorems that reduce the problem of computing the noise required for $\eps$-BDP to computing the noise for $\eps'$-DP, where $\eps > \eps'$. In other words, they determine the ``price'' (in terms of the privacy budget) of data correlations. Theorem 3 in \cite{zhao2017dependent} states that $\eps$-BDP $\impliedby \eps_3$-DP with: 
$$\eps_3 = \eps - 6 \ln \max_{x_{j+1}, x_j, x_j' \in \bits} \frac{\Pr[X_j = x_{j+1}|X_j = x_j]}{\Pr[X_j = x_{j+1}|X_j = x_j']}$$
So, in our simple case, $\eps_3 = \eps - 6 \ln \frac{1-\theta}{\theta}$. Using \autoref{noiseForDP} to solve for $\rho$ in terms of $\eps$:
\begin{align}
    \rho &\ge \frac{2(1-\theta)^6}{\theta^6 e^\eps + (1-\theta)^6} \label{noiseForThm3DDP}
\end{align}
However, with this, they cannot provide $\rho$ for $\eps$ below $6 \ln \frac{1-\theta}{\theta}$ (since $\eps_3$ can't be negative). To fix this, they provide a stronger theorem which constructs a piecewise linear function.
% To fix this they take the maximum of $\eps_3$ and $\eps/n$, since changing one record affects at most $n$ others (the entire dataset). However, their mechanism, which sets $\rho$ as above given some target $\eps$, has much higher noise for any given $\eps$, as seen in \autoref{fig:thm3_ddp_noise}.
\begin{proposition}[Adapted from \cite{zhao2017dependent}] Let $0 < \theta < 0.5$. Then, $\eps$-BDP $\impliedby \eps_6$-DP where:
\begin{align*}
    \epsilon_6 &= \max_{1\le t\le n/2}\frac{\epsilon -6 \ln \max_{i,j,k\in \{0,1\}}\frac{\Pr[X_t = k\mid X_0 = j]}{\Pr[X_t = k\mid X_0 = i]}}{2t-1} \\
    &= \max_{1\le t\le n/2}\frac{\epsilon -6 \ln \frac{1+(1-2\theta)^t}{1-(1-2\theta)^t}}{2t-1}
\end{align*}
\end{proposition}
However, there's no obvious closed form for $\rho$ in terms of $\eps$ and $\theta$ in this expression. Our work provides an approximate closed form and we have significantly better noise-privacy tradeoffs, as shown in \autoref{fig:thm3_ddp_noise}.
% \begin{proof}
%     Let $P = \begin{bmatrix} 1-\theta & \theta\\
%     \theta & 1-\theta
%     \end{bmatrix}$, and $e_0, e_1$ be the standard basis vectors.  For $j, k\in \{0,1\}$, the conditional probability $\Pr[X_t = k\mid X_0 = j]$ can be denoted as 
%     \begin{align*}
%         \Pr[X_t = k\mid X_0 &= j] = e^\top_j P^t e_k        \\
%         &= \frac{1}{2} e^\top_j \begin{bmatrix}
%        1+(1-2\theta)^t & 1-(1-2\theta)^t\\
%        1-(1-2\theta)^t & 1+(1-2\theta)^t
%        \end{bmatrix}e_k
%     \end{align*}
%        Therefore, 
%        \begin{align*}
%            \max_{i,j,k\in \{0,1\}}\frac{\Pr[X_t = k\mid X_0 = j]}{\Pr[X_t = k\mid X_0 = i]} &= \frac{1+(1-2\theta)^t}{1-(1-2\theta)^t} \\
%            &= \frac{2}{1-(1-2\theta)^t} - 1
%        \end{align*}
%        for $k = j \neq i$. Note that to maximize $\eps_6$ with respect to $t$, this quantity should be minimized (since it's always greater than 1, the log is always positive) by maximizing $t$. On the other hand, the denominator of $2t-1$ ought to be minimized by minimizing $t$. This tension means that there's no obvious closed form for $\eps_6$.
% \end{proof}

\subsection{Experimental Evaluation on Heartbeat Data}
\begin{figure*}[hbt!]
    \centering    
    \begin{subfigure}{.3\textwidth}
        \centering
        \begin{tikzpicture}
            \begin{axis}[
                legend style={font=\footnotesize},
                xlabel={Privacy Budget ($\eps$)},
                ylabel={Reconstruction Acc.},
                table/col sep=comma,
                xmin=1, xmax=4,
                ymin=0.5, ymax=1,
                xtick distance={0.5},
                ytick distance={0.1},
                legend pos=north west,
                ymajorgrids=true,
                grid style=dashed,
                width=0.9\linewidth,
            ]
            \addplot table [x=eps, y=viterbi_accuracy]{"data/synthetic_highcor_8e4lr.csv"};
            \addlegendentry{Viter.}
            \addplot table [x=eps, y=lstm_accuracy]{"data/synthetic_highcor_8e4lr.csv"};
            \addlegendentry{LSTM}  
            \addplot table [x=eps, y=bdp_bound]{"data/real_data_n5.csv"};
            \addlegendentry{BDP} 
            % \addplot+[mark=*, dashed] coordinates{(0.0, 0.622) (0.09, 0.622) (0.185, 0.622) (0.285, 0.622) (0.385, 0.622) (0.475, 0.622)};
            % \addlegendentry{DP Bound}
            \end{axis}
        \end{tikzpicture}
        \subcaption{High correlation, synthetic data}
    \end{subfigure}    
    \begin{subfigure}{.3\textwidth}
        \centering
        \begin{tikzpicture}
            \begin{axis}[
                legend style={font=\footnotesize},
                xlabel={Privacy Budget ($\eps$)},
                ylabel={Reconstruction Acc.},
                table/col sep=comma,
                xmin=1, xmax=4,
                ymin=0.5, ymax=1,
                xtick distance={0.5},
                ytick distance={0.1},
                legend pos=south east,
                ymajorgrids=true,
                grid style=dashed,
                width=0.9\linewidth,
            ]
            \addplot table [x=eps, y=viterbi_accuracy]{"data/real_data_n10.csv"};
            \addlegendentry{Viter.}
            \addplot table [x=eps, y=lstm_accuracy]{"data/real_data_n10.csv"};
            \addlegendentry{LSTM} 
            \addplot table [x=eps, y=bdp_bound]{"data/real_data_n5.csv"};
            \addlegendentry{BDP} 
            % \addplot+[mark=*, dashed] coordinates{(0.0, 0.622) (0.09, 0.622) (0.185, 0.622) (0.285, 0.622) (0.385, 0.622) (0.475, 0.622)};
            % \addlegendentry{DP Bound}
            \end{axis}
        \end{tikzpicture}
        \subcaption{High correlation, real data}
    \end{subfigure}    
    \begin{subfigure}{.3\textwidth}
        \centering
        \begin{tikzpicture}
            \begin{axis}[
                legend style={font=\footnotesize},
                xlabel={Privacy Budget ($\eps$)},
                ylabel={Reconstruction Acc.},
                table/col sep=comma,
                xmin=1, xmax=4,
                ymin=0.5, ymax=1,
                xtick distance={0.5},
                ytick distance={0.1},
                legend pos=south east,
                ymajorgrids=true,
                grid style=dashed,
                width=0.9\linewidth,
            ]
            \addplot table [x=eps, y=viterbi_accuracy]{"data/real_data_n5.csv"};
            \addlegendentry{Viter.}
            \addplot table [x=eps, y=lstm_accuracy]{"data/real_data_n5.csv"};
            \addlegendentry{LSTM} 
            \addplot table [x=eps, y=bdp_bound]{"data/real_data_n5.csv"};
            \addlegendentry{BDP} 
            \end{axis}
        \end{tikzpicture}
        \subcaption{Low correlation, real data}
    \end{subfigure}
    \caption{The reconstruction accuracy for both the Viterbi and the LSTM attackers is lower than our BDP bounds, on both synthetic (a) and real (b,c) data. There two real datasets correspond to a high correlation setting ($q = 0.0893, r = 0.1092, n = 26923$) and a low correlation setting ($q = 0.2384, r=0.3831, n=16859$). The synthetic data is chosen from the high correlation parameters.}
    \label{fig:real_data}
\end{figure*}
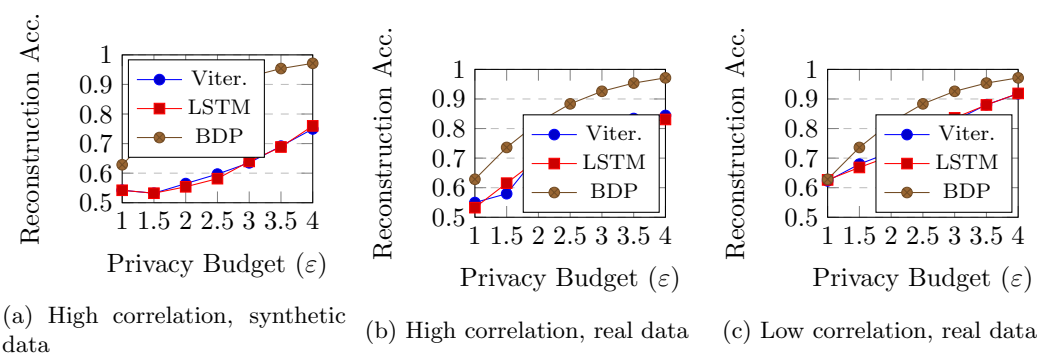

We also experiment with real world data. Real world data is not necessarily drawn from a true Markov chain -- there are often more complex correlation models at play, which may be unknown even to the data curator. Thus, a natural question is whether our guarantees extended meaningfully to data that is only approximately Markovian. In other words, can an adversary exploit any (hidden) underlying correlation in approximately Markovian data to violate the expected privacy guarantees of our mechanism? We provide some evidence to the contrary.

%We provide evidence to the contrary. 
We used a database containing heart rate data from several subjects while sleeping \cite{PENG1999101, goldberger2000physiobank}. We choose this setting because research suggests that heart rate data while sleeping exhibits many forms of correlation. During deep sleep, heart rates are relatively Markovian while during REM\footnote{Rapid eye movement} sleep, they have long-time correlations (e.g. $X_{10}$ influences $X_{100}$). And light sleep is a middle ground between these two extremes  \cite{penzel2003dynamics}. This is therefore an ideal setting to test whether (erronerously) modelling the data generation via Markov chains is problematic. The heart rate time series we use are around $15,000-25,000$ samples long, sampled at around 2Hz over 6 hours. We convert the heart rate data to binary data streams by clustering samples below and above the mean. 
%\jie{can we show a visual picture of both the heart beat data and the discretized one?}
To apply our algorithm, we first compute the empirical $q$ and $r$ from the binary data by observing the frequency of state transitions. Then, we sanitize the data via \autoref{mainMech}.

To see if a more sophisticated adversary can defeat the bounds of our mechanism, we construct an neural network adversary with a Long Short-Term Memory (LSTM)~\cite{lstm} architecture that tries to reconstruct the private database given the sanitized database. We collate the data with the input being a window of sanitized bit and the output being the hidden bit in the middle of this window (the window size is a hyperparameter). Our neural architecture is simple. We embed the inputs in a four dimensional space, feed it through the LSTM, and then apply a linear layer to the last LSTM state, using a sigmoid for our output. Despite being simple, choosing an appropriate window size enables our architecture to detect more complex correlation structures than a simple HMM. We contrast the LSTM attacker with a Viterbi attacker, which simply runs the Viterbi algorithm on the perturbed sequence to predict the private database. 

To calibrate our expectations, let the success probability be $p_s$. For $\eps$-BDP, we can provide a very similar bound as with $\eps$-DP:
\begin{align*}
    p_s \le \frac{e^\eps}{\max(\frac{q}{r},\frac{r}{q}) + e^\eps}
\end{align*}
The only difference from our earlier bound is due to the asymmetry of the Markov chain. Unlike DP however, we show that BDP does bound the correlation advantage.

As a sanity check, we first show that the LSTM attacker matches the Viterbi attacker on synthetic data constructed from a heavily correlated Markov chain. Then, we show that the LSTM attacker approximately matches the Viterbi attacker on two datasets -- the first is heavily correlated, with $q = 0.0893, r = 0.1092$, while the second has very asymmetric correlation, with $q = 0.2384, r = 0.3831$. We report the reconstruction accuracy of the Viterbi attacker on the entire database, and the validation accuracy of the LSTM attacker using 10-fold cross validation. In \autoref{fig:real_data}, we plot the reconstruction accuracy for both attackers as the privacy budget $\eps$ varies, in each setting. We also plot the theoretical bound that BDP guarantees, showing that both attackers are significantly below the BDP bound. Note that, unlike our DP experiments, the reconstruction accuracy \emph{decreases} for a fixed $\eps$ with more correlated data. This is because BDP takes the worst-case probability over all possible sanitizations, and the difference between the worst-case and the average-case broadens significantly as the data gets more correlated (this is why the SB attacker was extremely close to the DP bound in the previous section). 

Our results suggest that our mechanism is somewhat robust to data that's not strictly drawn from a Markov chain. In other words, despite sanitizing the data while assuming it is drawn from a Markov chain, the more sophisticated LSTM attacker was not able to get a significant advantage over the Viterbi attacker. This is despite the fact that the data exhibits more complex correlations \cite{penzel2003dynamics}, which an LSTM could learn. 

\section{Conclusion}
We investigated privacy definitions for correlated data, and found that Bayesian differential privacy (BDP) provides meaningful guarantees in the presence of data correlations. We described an optimal mechanism for achieving BDP over binary Markov chains. Our mechanism is both local, and can thus be implemented in a distributed manner, and non-interactive, outputting a sanitized database. Finally, in a series of experiments, we showed the harms of using DP in correlated settings, the improvement of our mechanism over the previous state-of-the-art, and of the robustness of our mechanism to approximately Markovian heart rate data.

We briefly suggest three possible directions to build upon this work. First, a basic goal would be to extend our model from binary Markov chains to any Markov chain with a finite state space. Similarly, our results can be extended from Markov chains to Markov random fields, which would better model data correlations found in social networks. Second, at the end of Section 3, we briefly discussed a correlated noise mechanism (which would lose local privacy), but found that it didn't lower the expected noise necessary. However, further investigation could reveal that correlated noise wins on other metrics, like the mutual information between the input and sanitized databases. Also, it may be possible to get better noise-privacy tradeoffs by dropping the some of the assumptions made for mathematical convenience. 

Lastly, at the end of Section 4, we discuss a more preliminary idea to reduce the amount of noise necessary with highly correlated data. BDP protects against the privacy loss when the mechanism produces the worst-case sanitization (i.e. when the all zero database is output), but the difference between this worst-case and the average-case can be quite large with sanitized data, as our experiments show. So one potential idea is to prevent the mechanism from outputting sanitizations close to the worst case; this would not provide local privacy, but may achieve better noise-privacy tradeoffs.

\appendix
\section{Detailed Proofs}
\subsection{Proof of \autoref{lm:lrMaxZ}}\label{appendix1}
\begin{proof}
The following two claims yield the lemma: 
\begin{enumerate}
    \item The ratio of forward probability, $\alpha_i(0; \mathbf{z})/\alpha_i(1; \mathbf{z})$ is maximized when $z_{1:t} = \mathbf{0}$. 
    \item The ratio of backward probability, $\beta_i(0; \mathbf{z})/\beta_i(1; \mathbf{z})$ is maximized when $z_{t+1:n} = \mathbf{0}$.
\end{enumerate}

For the first claim, we go by induction on $i$.  For the base case, $i = 1$, because $\rho_0$ and $\rho_1$ are smaller than $1/2$:
\begin{align*}
    \frac{\alpha_1(0;1)}{\alpha_1(1;1)} =& \frac{\Pr[Z_1 = 1\mid X_1 = 0]}{\Pr[Z_1 = 1\mid X_1 = 1]}=\frac{\rho_0}{1-\rho_1}\le 1\\
    \le & \frac{1-\rho_0}{\rho_1}  = \frac{\Pr[Z_1 = 0\mid X_1 = 0]}{\Pr[Z_1 = 0\mid X_1 = 1]} = \frac{\alpha_1(0;0)}{\alpha_1(1;0)}.
\end{align*}
Now, suppose the claim is true for $i \ge 1$. Using basic probability theory, $\alpha_{i+1}(x)$ can be recursively computed from $\alpha_i$:
\begin{align*}
    \alpha_{i+1}(x;\mathbf{z}) = \alpha_i(0)\,B_{x, z_{i+1}}\, P_{x, 0}\cdot +\alpha_i(1)\,B_{x, z_{i+1}}\, P_{x, 1}
\end{align*}
Where $P$ and $B$ are the transition and noise matrices, respectively.
Therefore the $\alpha$ ratio is: 
\begin{align*}
    \frac{\alpha_{i+1}(0)}{\alpha_{i+1}(1)} =& \frac{\alpha_i(0)\,B_{0, z_{i+1}}\, P_{0, 0}+\alpha_i(1)\,B_{0, z_{i+1}}\, P_{0, 1}}{\alpha_i(0)\,B_{1, z_{i+1}}\, P_{1, 0}+\alpha_i(1)\,B_{1, z_{i+1}}\, P_{1, 1}}\\
    =& \frac{B_{0, z_{i+1}}}{B_{1, z_{i+1}}}\cdot \frac{\frac{\alpha_i(0)}{\alpha_i(1)} (1-q)+q}{\frac{\alpha_i(0)}{\alpha_i(1)} r +\, (1- r)}
\end{align*}
To maximize this, note that the first term only depends on $z_{i+1}$ and the second term only depends on the value of $\mathbf{z}_{1:i}$.  Hence, we can maximize these two terms separately.
The first term is identical to the base case, and is thus maximized by $z_{i+1} = 0$.
% For the first term, because $\rho_0$ and $\rho_1$ are less than $1/2$, we have $B_{0,0}/B_{1,0} = 1-\rho_0/\rho_1\ge 1\ge \rho_0/(1-\rho_1) = B_{0,1}/B_{1,1}$, so $z_{i+1} = 0$ maximizes the first term.
% We also claim the second term is increasing in $\frac{\alpha_i(0)}{\alpha_i(1)}$. By the quotient rule, the derivative is positive if: notice, that $1-q > r$, so the top is increasing faster than the bottom. 
For the second term, one can simply take the derivative with respect to the $\alpha$ ratio, and since $(1-q)(1-r)>qr$, this term is is increasing in $\frac{\alpha_i(0)}{\alpha_i(1)}$.
Therefore, by our inductive hypothesis, the second term is maximized when $\mathbf{z}_{1:t} = \mathbf{0}$.  Combining both terms, we've shown that the ratio $\frac{\alpha_{i+1}(0)}{\alpha_{i+1}(1)}$ is maximized when $\mathbf{z}_{1:t+1} = \mathbf{0}$. This completes the proof of our first claim.

The second claim is almost exactly symmetrical, except we must go by induction from $i = n$ downwards, since the $\beta$ recurrence is:
\begin{equation*}
    \beta_i(x; \mathbf{z}) = \beta_{i+1}(0) B_{0, z_{i+1}} P_{x,0} + \beta_{i+1}(1)B_{1,z_{i+1}}P_{x,1}
\end{equation*}
The remaining steps are symmetrical.
\end{proof}

\subsection{Proof of \autoref{lm:closedForm}}\label{appendix2}
\begin{proof}
Throughout this proof, let $\mathbf{z} = \mathbf{0}$ and let $i \in [n]$ be arbitrary.
From our earlier derivations, the recurrence for $\alpha$ is:
$$\begin{bmatrix}
\alpha_{i}(0)\\
\alpha_{i}(1)
\end{bmatrix} = \begin{bmatrix}
(1-\rho_0)(1-q) & (1-\rho_0)q\\
\rho_1 r & \rho_1(1-r)
\end{bmatrix}\begin{bmatrix}
\alpha_{t-1}(0)\\
\alpha_{t-1}(1)
\end{bmatrix}$$
With base case $\alpha_0(x) = 1$. Now, solving this recurrence for arbitrary $i$ just involves taking powers of the $2 \times 2$ matrix:
$$\begin{bmatrix}
\alpha_{i}(0)\\
\alpha_{i}(1)
\end{bmatrix} = \begin{bmatrix}
(1-\rho_0)(1-q) & (1-\rho_0)q\\
\rho_1 r & \rho_1(1-r)
\end{bmatrix}^t\begin{bmatrix}
1\\
1
\end{bmatrix}$$
To solve this, we diagonalize the matrix by finding the eigenvalues and eigenvectors. Let $V$'s columns be the eigenvectors and $\Lambda$ be the diagonal matrix of eigenvalues. Then:
$$\begin{bmatrix}
    \alpha_{i}(0)\\
    \alpha_{i}(1)
    \end{bmatrix} = V \Lambda^i V^{-1}\begin{bmatrix}
    1\\
    1
\end{bmatrix}$$
First, we compute the eigenvalues and eigenvectors. Let $x = (1-\rho_0)(1-q)$ and $y = \rho_1(1-r)$. Then:
\begin{align*}
    \lambda_1 &= 0.5\Big(x + y +\sqrt{(x+y)^2 - 4\rho_1(1 - q - r)(1 -\rho_0)}\Big) \\
    \lambda_2 &= 0.5\Big(x + y -\sqrt{(x+y)^2 - 4\rho_1(1 - q - r)(1 -\rho_0)}\Big) \\ 
    v_1 &= \begin{bmatrix}
        x - y + \sqrt{(x-y)^2 +4qr(1-\rho_0)\rho_1} \\
        2r\rho_1
    \end{bmatrix} \eqdef \begin{bmatrix}
        a \\
        c
        \end{bmatrix} \\
    v_2 &= \begin{bmatrix}
        x - y - \sqrt{(x-y)^2 +4qr(1-\rho_0)\rho_1}  \\
        2r\rho_1
    \end{bmatrix} \eqdef \begin{bmatrix}
        b \\
        c
        \end{bmatrix} 
\end{align*}
Putting this back into the recurrence:
\begin{align*}
    \begin{bmatrix}\alpha_{i}(0)\\ \alpha_{i}(1)\end{bmatrix} &= \begin{bmatrix}ac\lambda_1^{i}-bc\lambda_2^{i} & ab(\lambda_2^{i} - \lambda_1^{i})\\ c^2(\lambda_1^{i} - \lambda_2^{i})& ac\lambda_2^{i} - bc\lambda_1^{i} \end{bmatrix} \cdot \begin{bmatrix} 1 \\ 1 \end{bmatrix} \\
    &=\begin{bmatrix}a (c - b)\lambda_1^{i} + b (a- c)\lambda_2^{i} 
    \\   c (c-b)\lambda_1^{i} + c (a- c)\lambda_2^{i}\end{bmatrix}
\end{align*}
Now we can compute the desired quantity:
\begin{equation}\label{eq:ratio_closed_alpha}
    \frac{\alpha_{i}(0)}{\alpha_{i}(1)} = \frac{a (c - b)\lambda_1^{i} + b (a- c)\lambda_2^{i}}{c (c-b)\lambda_1^{i} + c (a- c)\lambda_2^{i}}\rightarrow \frac{a}{c} \text{ as } i\to \infty
\end{equation}
We now turn our attention to the $\beta$ recurrence and perform very similar steps:
$$\begin{bmatrix}
\beta_{i}(0)\\
\beta_{i}(1)
\end{bmatrix} = \begin{bmatrix}
(1-\rho_0)(1-q) & \rho_1 q\\
(1-\rho_0) r & \rho_1(1-r)
\end{bmatrix}^{n-i}\begin{bmatrix}
1\\
1
\end{bmatrix}$$
The eigenvalues of this matrix are the same, and for the eigenvectors, the only change is that $c$ changes to $d \eqdef 2r(1-\rho_0)$. Thus, our final ratio is:
\begin{equation}\label{eq:ratio_closed_beta}
    \frac{\beta_{i}(0)}{\beta_{i}(1)} = \frac{a (d - b)\lambda_1^{n-i} + b (a- d)\lambda_2^{n-i}}{d (d-b)\lambda_1^{n-i} + d (a- d)\lambda_2^{n-i}}\rightarrow \frac{a}{d} \text{ as } n - i\to \infty
\end{equation}
We now claim that the $\alpha$ and $\beta$ ratios from \autoref{eq:ratio_closed_alpha} and \autoref{eq:ratio_closed_beta} are bounded above by their limit, i.e. that:
\begin{align*}
    \frac{\alpha_{i}(0)}{\alpha_{i}(1)} \le \frac{a}{c} \text{ and }
    \frac{\beta_{i}(0)}{\beta_{i}(1)} \le \frac{a}{d}
\end{align*}
To do so, we first claim that $a \ge c, d \ge 0 \ge b$. We omit the proof of this claim in this version.
Now, we can rewrite \autoref{eq:ratio_closed_alpha} as:
\begin{align*}
    &\frac{\alpha_{i}(0)}{\alpha_{i}(1)} = \frac{a + b \frac{a- c}{c - b}\left(\frac{\lambda_2}{\lambda_1}\right)^{i}}{c + c \frac{a- c}{c - b}\left(\frac{\lambda_2}{\lambda_1}\right)^{i}} \eqdef \frac{a + b'}{c + c'} 
\end{align*}
Now, notice that $a \ge a + b'$, since $b' \le 0$ and $c \le c + c'$ since $c' \ge 0$. Thus, $a/c$ is an upper bound for the $\alpha$ ratio.
We can do the exact same thing for the $\beta$ ratio, since we also know that $a \ge d$, concluding the proof.
\end{proof}

\subsection{Proof of \autoref{lm:lrMaxI}}\label{appendix3}
\begin{proof}
Recall from earlier that: 
\begin{align*}
    \frac{\Pr[\mathbf{Z} = \mathbf{z}\mid X_i = 0]}{\Pr[\mathbf{Z}= \mathbf{z}\mid X_i = 1]} = \frac{\alpha_i(0)}{\alpha_i(1)}\cdot \frac{\beta_i(0)}{\beta_i(1)}
\end{align*}
Let $\sigma = \lambda_2/\lambda_1$, $A = a (c - b)$, $A' = a (d - b)$, $B = -b (a- c)$, $B' = - b(a - d)$, $C = c (c-b)$, $C' = d(d - b)$, $D = c (a- c)$, and $D' = d(a - d)$. Now, from \autoref{eq:ratio_closed_alpha} and \autoref{eq:ratio_closed_beta}, we know that:
\begin{align*}
    \frac{\alpha_{j}(0)}{\alpha_{j}(1)} = \frac{a (c - b)\lambda_1^{i} + b (a- c)\lambda_2^{i}}{c (c-b)\lambda_1^{i} + c (a- c)\lambda_2^{i}} = \frac{A-B\sigma^i}{C+D\sigma^i} \\
    \frac{\beta_{i}(0)}{\beta_{i}(1)} = \frac{a (d - b)\lambda_1^{n-i} + b (a- d)\lambda_2^{n-i}}{d (d-b)\lambda_1^{n-i} + d (a- d)\lambda_2^{n-i}} = \frac{A'-B'\sigma^{n-i}}{C'+D'\sigma^{n-i}}\\
    \frac{\Pr[\mathbf{z}\mid X_i = 0]}{\Pr[\mathbf{z}\mid X_i = 1]} = \frac{AA' -AB'\sigma^{n-i} - A'B\sigma^i + BB'\sigma^n}{CC' + CD'\sigma^{n-i} + C'D \sigma^i + DD'\sigma^n}
\end{align*}
Since only the two middle terms (in the numerator and denominator) involve $i$, we focus our attention there. We want to combine these terms, so notice that:
\begin{align*}
    AB' = \frac{(a-d)(c-b)}{(a-c)(d-b)}\cdot A'B \\
    CD' = \frac{(a-d)(c-b)}{(a-c)(d-b)}\cdot C'D
\end{align*}
Let $\gamma = \frac{(a-d)(c-b)}{(a-c)(d-b)}$. Then, we can rewrite the likelihood ratio as:
\begin{align*}
    \frac{\Pr[\mathbf{Z} = \mathbf{z}\mid X_i = 0]}{\Pr[\mathbf{Z}= \mathbf{z}\mid X_i = 1]} &= \frac{AA' -A'B(\gamma\sigma^{n-i} + \sigma^i) + BB'\sigma^n}{CC' + C'D(\gamma\sigma^{n-i} + \sigma^i) + DD'\sigma^n}
\end{align*} 
Now, since $A', B, C', \text{ and } D$ are all positive (by our previous claims), by minimizing $\gamma\sigma^{n-i} + \sigma^i$, we maximize the ratio (all other terms are independent of $i$). Since $\gamma$ is also positive, we can apply the AM-GM inequality, which says that $0.5(\gamma\sigma^{n-i}+\sigma^i) \ge \sqrt{\gamma\sigma^n}$. Note that the right hand side is a constant with respect to $i$. The AM-GM inequality also tells us that we get equality iff $\sigma^{i} = \gamma\sigma^{n-i}$, so doing so minimizes $\gamma\sigma^{n-i}+\sigma^i$. Taking logs:
\begin{align*}
    i \log \sigma &= \log \gamma + (n-i) \log \sigma\\
    i &= \frac{\log \gamma + n\log \sigma}{2\log \sigma} = \frac{n}{2} + \frac{\log \gamma}{2\log\sigma} = \frac{n}{2} \pm \mathcal{O}(1)
\end{align*}
We now argue that this shows that our theorem bound is tight up to negligible factors.
Recall that the exact $\alpha$ ratio is:
\begin{align*}
    \frac{\alpha_{i}(0)}{\alpha_{i}(1)} = \frac{a + b \frac{a- c}{c - b}\left(\frac{\lambda_2}{\lambda_1}\right)^{i}}{c + c \frac{a- c}{c - b}\left(\frac{\lambda_2}{\lambda_1}\right)^{i}}
\end{align*}
Notice that because $\sigma = \lambda_2/\lambda_1 < 1$, the $\alpha$ ratio approaches $a/c$ exponentially quickly as $i$ increases; similarly, the $\beta$ ratio approaches $a/d$ exponentially quickly as $n-i$ increases. Since we know that the maximum index is on the order of $n/2$, we know that both ratios approach their limit up to some inverse exponential factors. (This also means that we can ignore the constant with again some inverse exponential loss.) In other words, with $z = \mathbf{0}$ and $i = n/2$:
\begin{align*}
    \frac{\Pr[\mathbf{Z} = \mathbf{0}\mid X_{n / 2} = 0]}{\Pr[\mathbf{Z}= \mathbf{0}\mid X_{n / 2} = 1]} = \frac{a^2 + o(\sigma^{n / 2})}{cd + o(\sigma^{n / 2})}
\end{align*}
\end{proof}
\subsection{Proof of \autoref{thm:bdpProof}}\label{appendix4}
As mentioned in the body, we present the proof for a symmetric Markov chain here for simplicity, but the result holds for the asymmetric case as well. Recall that we assume $0 < \theta < 0.5$, where $\theta$ is the transition parameter (the probability that the state changes). 
\begin{proof}
    It's obvious that the BDPL will always be maximized with $x_i, \mathbf{x}_{K}, \mathbf{z} = \mathbf{0}$, since the Markov chain is lazy and the mechanism is more likely to keep a bit than flip it (this can be shown formally via induction, similar to \autoref{lm:lrMaxZ}). So, let $\mathbf{z}, x_i, \mathbf{x}_K$ all be set to zero here. The privacy losses can be written as:
    \begin{align*}
        BDPL(A) = \frac{\Pr[z_i|x_i]}{\Pr[z_i|\bar{x}_i]} \cdot \frac{\Pr[\mathbf{z}_U|x_i, x_j, \mathbf{x}_{K'}]} {\Pr[\mathbf{z}_U|\bar{x}_i, x_j, \mathbf{x}_{K'}]} \\
        BDPL(A') = \frac{\Pr[z_i|x_i]}{\Pr[z_i|\bar{x}_i]} \cdot \frac{\Pr[\mathbf{z}_U, z_j|x_i, \mathbf{x}_{K'}]} {\Pr[\mathbf{z}_U, z_j|\bar{x}_i, \mathbf{x}_{K'}]}
    \end{align*}
    Notice that if $j$ is not adjacent to $U$:
    \begin{align*}
        BDPL(A') = \frac{\Pr[z_i|x_i]}{\Pr[z_i|\bar{x}_i]}\cdot \frac{\Pr[\mathbf{z}_U|x_i, \mathbf{x}_{K'}]} {\Pr[\mathbf{z}_U|\bar{x}_i, \mathbf{x}_{K'}]} \cdot \frac{\Pr[z_j|x_i, \mathbf{x}_{K'}]} {\Pr[z_j|\bar{x}_i, \mathbf{x}_{K'}]}
    \end{align*}
    The first two fractions are simply $BDPL(A)$ (since by conditional independence $\Pr[\mathbf{x}_U|x_i, x_j, \mathbf{x}_{K'}] = \Pr[\mathbf{x}_U|x_i, \mathbf{x}_{K'}]$). For the last fraction, notice first that if $j$ is not adjacent to $i$, this last fraction is simply one, since we can ignore $x_i$ via conditional independence. Otherwise, we use the following lemma.
    \begin{lemma}
        Suppose $j$ is $i-1$ or $i+1$, and let $j'$ be $i-2$ or $i+2$ respectively. Then:
        \begin{align*}
            \frac{\Pr[z_j|x_i, x_{j'}]} {\Pr[z_j|\bar{x}_i, x_{j'}]} = 2  \frac{(1-\rho)(1-\theta)^2 + \rho\theta^2}{(1-\theta)^2 + \theta^2}
        \end{align*}
    \end{lemma}
    The lemma involves just a basic calculation, and so we omit the proof.
    Since $j$ is not adjacent to $U$, we know that $j$ is between $i$ and a known tuple; thus the lemma applies and the BDPL increases. 

    Now, consider the case where $j$ is adjacent to $U$. Without loss of generality, assume $j$ is to the left of $U$, and write $U$ as $U = U_L = \{i-k, \dots, i-1\} \cup U_R \{i+1, \dots, r\}$, the union of the portion to the left and right of $x_i$ (so $j = i-k-1$). Note that $U_L$ or $U_R$ could be empty. By conditional independence:
    \begin{align*}
        BDPL(A) &= \frac{\Pr[z_i|x_i]}{\Pr[z_i|\bar{x}_i]} \cdot \frac{\Pr[\mathbf{z}_{U_L}|x_i, x_j]} {\Pr[\mathbf{z}_{U_L}|\bar{x}_i, x_j]} \cdot \frac{\Pr[\mathbf{z}_{U_R}|x_i, x_r]} {\Pr[\mathbf{z}_{U_R}|\bar{x}_i, x_r]}  \\
        BDPL(A') &= \frac{\Pr[z_i|x_i]}{\Pr[z_i|\bar{x}_i]} \cdot \frac{\Pr[\mathbf{z}_{U_L}, z_j|x_i, x_{j-1}]} {\Pr[\mathbf{z}_{U_L}, z_j|\bar{x}_i, x_{j-1}]} \cdot \frac{\Pr[\mathbf{z}_{U_R}|x_i, x_r]} {\Pr[\mathbf{z}_{U_R}|\bar{x}_i, x_r]} 
    \end{align*}
    For comparative purposes, we ignore the first and last terms in the loss expressions. We also substitute $i-k-1$ for $j$:
    \begin{align*}
        BDPL(A) &\propto \frac{ \Pr[\mathbf{z}_{i-k:i-1}|x_i, x_{i-k-1}]} {\Pr[\mathbf{z}_{i-k:i-1}|\bar{x}_i, x_{i-k-1}]} \\
        BDPL(A') &\propto \frac{\Pr[\mathbf{z}_{i-k-1:i-1}|x_i, x_{i-k-2}]} {\Pr[\mathbf{z}_{i-k-1:i-1}|\bar{x}_i, x_{i-k-2}]}
    \end{align*}
    Now, let $\gamma_k(x,y) = \Pr[z_{i-k:i-1}, X_i = x, X_{i-k-1} = y]$. Then:
    \begin{align*}
        BDPL(A) &\propto \frac{\gamma_k(x_i, x_{i-k-1})/\Pr[x_i, x_{i-k-1}]}
        {\gamma_k(x_i, x_{i-k-1})/\Pr[x_i, x_{i-k-1}]} \\
        &= \frac{\gamma_k(x_i, x_{i-k-1})}
        {\gamma_k(\bar{x}_i, x_{i-k-1})} \cdot \frac{\Pr[\bar{x}_i, x_{i-k-1}]}{\Pr[x_i, x_{i-k-1}]}\\
        BDPL(A') &\propto
        \frac{\gamma_{k+1}(x_i, x_{i-k-2})}{\gamma_{k+1}(\bar{x}_i, x_{i-k-2})}  \cdot 
        \frac{\Pr[\bar{x}_i, x_{i-k-2}]}{\Pr[x_i, x_{i-k-2}]}
    \end{align*}
    Since $\mathbf{z} = \mathbf{0}$, we can write $\gamma$ recursively as: 
    \begin{align*}
        \gamma_k(x,0) &= (1-\theta)(1-\rho) \gamma_{k-1}(x,0) + \theta \rho \gamma_{k-1}(x,1) \\
        \gamma_k(x,1) &= \theta (1-\rho) \gamma_{k-1}(x,0) + (1-\theta)\rho \gamma_{k-1}(x,1)
    \end{align*}
    For $x \in \bits$. This exactly matches the $\beta$ recurrence from our main proof, except for the base case. When $k = 0$, $\gamma_0(0,0), \gamma_0(1,1) = 0.5(1-\theta)$ and $\gamma_0(0,1), \gamma_0(1,0) = 0.5\theta$. The eigenvectors and eigenvalues are:
    \begin{align*}
        \lambda_1 &= 0.5(\sqrt{(1 - \theta)^2 - (1 - 2\theta)(1 - 2\rho)^2} + (1 -\theta)) \\
        \lambda_2 &= 0.5(-\sqrt{(1 - \theta)^2 - (1 - 2\theta)(1 - 2\rho)^2} + (1 -\theta)) \\
        v_1 &= \begin{bmatrix} \sqrt{\theta^2 + (1 - 2\theta)(1 - 2\rho)^2} + (1 -\theta)(1 - 2\rho)\\ 2\theta(1 -\rho)
        \end{bmatrix} \eqdef \begin{bmatrix}
            a \\
            c
            \end{bmatrix} \\
        v_2 &= \begin{bmatrix} -\sqrt{\theta^2 + (1 - 2\theta)(1 - 2\rho)^2} + (1 -\theta)(1 - 2\rho)\\ 2\theta(1 -\rho)
        \end{bmatrix} \eqdef \begin{bmatrix}
            b \\
            c
            \end{bmatrix}\\
        V&=\begin{bmatrix}v_1 & v_2 \end{bmatrix} = \begin{bmatrix}a & b\\ c& c \end{bmatrix}, V^{-1} = \begin{bmatrix}c & -b\\ -c& a \end{bmatrix}
    \end{align*}
    Thus, we get the following closed forms:
    \begin{align*}
        \gamma_k(0,0) &= (1-\theta)(ac\lambda_1^k - bc \lambda_2^k) + \theta ab(\lambda_2^k - \lambda_1^k) \\
        &= \lambda_1^k a \big((1-\theta)c - \theta b\big) + \lambda_2^k b \big(\theta a - (1-\theta) c\big) \\
        \gamma_k(1,0) &= \theta(ac\lambda_1^k - bc \lambda_2^k) + (1-\theta) ab(\lambda_2^k - \lambda_1^k) \\
        &= \lambda_1^k a \big(\theta c - (1-\theta) b\big) + \lambda_2^k b \big((1-\theta) a - \theta c\big)
    \end{align*}
    Turning attention back to the BDPL, let $l = \lambda_2/\lambda_1$; note $l < 1$. We define the following functions:
    \begin{align*}
        f(k) &\eqdef \frac{\Pr[X_i = 1, X_{i-k-1} = 0]}{\Pr[X_i = 0, X_{i-k-1} = 0]} = \dfrac{1-(1-2\theta)^{k+1}}{1+(1-2\theta)^{k+1}} \\
        g(k) &\eqdef \frac{\gamma_k(0,0)}{\gamma_k(0,1)} = \dfrac{a(c(1-\theta)-b\theta)+bl^k(a\theta-c(1-\theta))}{a(c\theta-b(1-\theta))+bl^k(a(1-\theta)-c\theta)}
    \end{align*}
    Thus, $BDPL(A) \propto f(k)g(k) \text{ and } BDPL(A') \propto f(k+1)g(k+1)$. From this closed form, we confirmed numerically that $h(k) \eqdef f(k)g(k)$ is increasing in $k$. This completes the proof.
\end{proof}
\bibliographystyle{abbrvnat}
\bibliography{refs}

\end{document}